\documentclass[10pt,a4paper,aps,showpacs,amssymb,pra,floatfix,footinbib,twocolumn,reprint,showkeys,superscriptaddress]{revtex4-1}
\usepackage[tbtags,intlimits]{amsmath}

\usepackage{slashed}
\usepackage{xcolor}
\usepackage{hyperref}
\usepackage[squaren]{SIunits}

\usepackage[pdftex]{graphicx}
\usepackage{grffile} 

\usepackage{bbold}

\renewcommand{\mathbf}{\boldsymbol}
\renewcommand{\d}{  {\mathrm d}   }

\begin{document}

\title{Laser assisted Compton scattering of X-ray photons}
\author{D.~Seipt}
\email{d.seipt@gsi.de}

\affiliation{Helmholtz-Institut Jena, Fr{\"o}belstieg 3, 07743 Jena, Germany}
\affiliation{Helmholtz-Zentrum Dresden-Rossendorf, P.O.~Box 510119, 01314 Dresden, Germany}

\author{B.~K{\"a}mpfer}
\affiliation{Helmholtz-Zentrum Dresden-Rossendorf, P.O.~Box 510119, 01314 Dresden, Germany}
\affiliation{TU Dresden, Institut f{\"u}r Theoretische Physik, 01062 Dresden, Germany}


\begin{abstract}
The Compton scattering of X-ray photons, assisted by a
short intense optical laser pulse, is discussed.
The differential scattering cross section reveals the interesting feature
that the main Klein-Nishina line
is accompanied by a series of side-lines
forming a broad plateau where
up to ${\cal O} (10^3)$ 
laser photons participate simultaneously in a single
scattering event.
An analytic formula for the width of the plateau is given.
Due to the non-linear mixing of X-ray and laser photons
a frequency dependent rotation of the polarization
of the final state X-ray photons
relative to the scattering plane emerges.
A consistent description of the scattering process with short laser pulses
requires to work with X-ray pulses.
An experimental investigation can be accomplished, 
e.g., at  LCLS or the European XFEL in the near future.
\end{abstract}

\maketitle

\section{Introduction}

Compton scattering \cite{Compton:PR1923}, i.e.~the
scattering of X or $\gamma$ rays off
free electrons is one of the fundamental interaction processes
of photons with charged particles. 
A particular feature is the dependence of the frequency $\omega_{\rm KN}'$ of the
scattered photon on the angle $\vartheta$,
$\omega_{\rm KN}' = \omega_X/(1 + \frac{\omega_X}{m} [1 - \cos \vartheta])$,
in the initial rest frame of the scattering particle with mass $m$.
For linearly polarized
X-rays
the scattered photon
polarization direction is the same as for electric dipole radiation.
These properties become modified in laser assisted Compton scattering
$X + L + e \to e' + X'$, where we suppose alignment of the X-ray beam
($X$, frequency $\omega_X\sim \mathcal O(\kilo\electronvolt)$)
and an intense optical laser pulse
($L$, frequency $\omega_L\sim \mathcal O(\electronvolt)$).
The frequency $\omega'$ of the scattered photon $X'$ in the initial rest frame of the electron $e$ reads (with $\hbar=c=1$)
\begin{align}
\omega' (\ell,\vartheta) & 
=  \frac{ \omega_X + \ell \omega_L}{ 1 + \frac{ \omega_X + \ell\omega_L}{m} (1-\cos \vartheta) } \,. \label{eq:frequency}
\end{align}
That means a non-linear frequency mixing occurs with
$\ell$ parametrizing the amount of energy and momentum absorbed from
the laser field in the scattering process.
The quantity $\ell$ can
be related to the number of involved laser photons, in particular in
the limit of infinite monochromatic plane waves, where the values of $\ell$ become
discrete $\ell_N$, with integers $N$ referring to the number of exchanged laser photons
(see Appendix \ref{app:IPW} for details).
The value of $\ell$ may be positive or negative,
leading to the formation of side-bands in the energy spectrum \cite{Ehlotzky:JPB1987}.
A similar effect has been observed also for laser assisted atomic processes \cite{Weingartshofer:PRL1977,Taieb:JPCS2008}.
Obviously, $\ell=0$ recovers the laser-free scattering
of an X-ray photon with known Klein-Nishina (KN) kinematics.
A large frequency ratio $\varkappa = \omega_X/\omega_L$ leads to a strong
dynamical enhancement of non-linear multi-laser photon effects 
such that many side-lines form a broad plateau
reaching to $|\ell| \gg 1$.
Although the possibility of a strong enhancement was recognized in \cite{Ehlotzky:JPB1987}, the
shape of the spectrum in frequency space and the cut-off values of $\ell$,
in particular their angular dependencies,
have never been calculated precisely to the best of our knowledge. 
This gap will be filled in this paper, where we calculate the frequency spectrum
and provide a formula for the angular dependence of the cut-off energies of the side-band plateau.

Observing the plateau of side-lines within a certain frequency interval determined by
cut-off values with special angular dependencies
is a clear experimental signal for
laser assisted Compton scattering of X-rays, i.e.\ a
new possibility to observe ${\cal O} (10^3)$ multi-photon effects in 
strong-field QED. In addition, as we shall show below, the polarization
properties of $X'$ represent a new feature.
The non-linear frequency mixing leads to
a frequency dependent rotation of the polarization of the final state photons.
This rotation does not affect the main KN line ($\ell = 0$) and is useful to identify
the side-bands in an experiment.

The basic scattering process can be understood qualitatively in a classical
picture, where the ``slow'' electron motion due to the laser is described
classically.
For high-intensity laser fields of the order of
$\unit{10^{18}}{\watt\per\centi\metre^2}$ the motion of electrons
becomes relativistic and non-linear, resembling
a figure-8 motion
with a velocity component in the laser beam direction
\cite{Schappert:PRD1970,DiPiazza:RevModPhys2012}, superimposed
to the usual transverse motion.
In this picture, laser assisted Compton scattering corresponds
to the scattering of X-rays off accelerated charges,
and the broadening of the KN line occurs due to
a time dependent Doppler shift induced by the
figure-8 motion.

In our approach we fully take into
account the finite lengths of both the X-ray and laser pulses,
going beyond infinite monochromatic plane wave approximation
\cite{Oleinik:JETP1968,Gush:PRD1975,
Akhiezer:JETP1985,Ehlotzky:JPB1987,Puntajer:PRA1989,Ehlotzky:JPB1989,Zhukovskii:JETP1973,book:Baier}
or the limit of long laser pulses and infinite X-ray waves \cite{Nedoreshta:LasPhys2013}.
Pulse shape effects have been recently proved to have a significant impact of
the scattering spectra in
non-linear Compton scattering
\cite{Narozhnyi:JETP83,Boca:PRA2009,Seipt:PRA2010,Seipt:PRA2011,
Mackenroth:PRA2011,Krajewska:PRA2012,Seipt:PRA2013},
pair production \cite{Heinzl:PLB2010,Nousch:PLB2012,Krajewska:PRA2013,Kohlfurst:PRD2013}
and other strong field QED processes \cite{Ilderton:PRA2011,DiPiazza:RevModPhys2012}.

Taking into account the finite pulse length is necessary as we
envisage a specific experimental set-up involving a petawatt class optical laser in combination with an X-ray free electron laser, which both produce short femtosecond light pulses.
In some previous papers, relations for the cross section were calculated either
for small numbers of exchanged laser photons $N$ \cite{Oleinik:JETP1968,Gush:PRD1975,Akhiezer:JETP1985},
or they were limited to non-relativistic intensities \cite{Ehlotzky:JPB1987},
or were restricted to the discussion of energy-integrated
angular distributions \cite{Puntajer:PRA1989};
the partial cross
sections have been calculated for arbitrary values of $N$
in terms of generalized Bessel functions, e.g.~in \cite{Ehlotzky:JPB1989}.

\begin{figure}
\center~\includegraphics[width=\columnwidth]{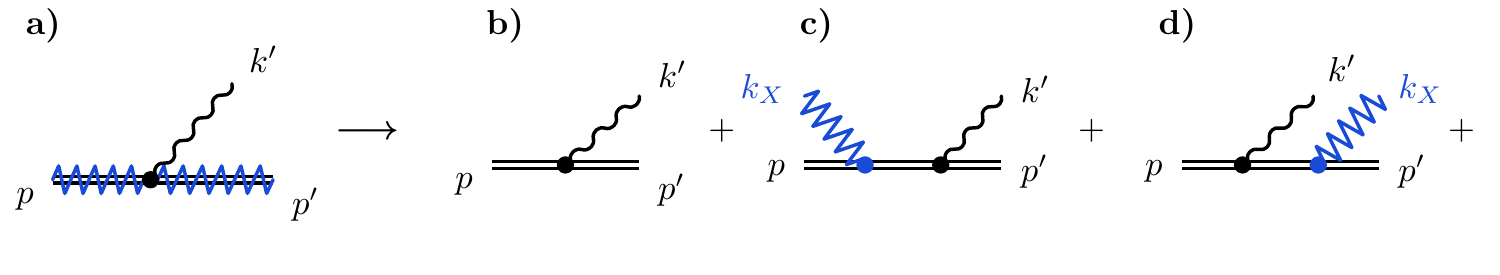}
\caption{(Color online)
Expansion of the Feynman diagram (a) for photon emission off
a dressed electron
(double zig-zag line: electron in the combined laser and X-ray field,
plain double line: electron in the laser background,
wavy line: photon, and zig-zag line: X-ray) for a weak X-ray field
in powers of $a_X$,
yielding spontaneous emission (b), laser assisted Compton scattering (c)
and induced emission (d).
}
\label{feyn:lacs}
\end{figure}

Our paper is organised as follows: In Section \ref{sect:matrix} we formulate the matrix element
for photon emission in the combined X-ray and laser fields and discuss the physical significance of the
various terms occurring in a weak field expansion in the X-ray field.
In Section \ref{sect:crosssection} the leading order contribution in the X-ray field is evaluated
and used to calculate an expression for the cross section for laser assisted Compton scattering.
The essential properties of the photon spectrum are discussed
in Section \ref{sect:spectrum} where we present results for the differential cross section
as well as the final photon polarization for a representative set of parameters.
High order multi-photon effects are quantified
via an analytic formula for the cut-off values of the frequency spectrum.
We discuss and summarize our results in Section \ref{sect:summary} where we also
address some aspects of an experimental realization. Appendix \ref{app:IPW} provides
a brief discussion on the limit of infinitely long plane waves.

\section{Matrix Element}
\label{sect:matrix}

We describe the photon beams as a classical
plane wave
background field with four-vector potential
$A^\mu = A^\mu_X(\varkappa\phi) + A^\mu_L(\phi)$,
defining $\phi = k_L \cdot x\equiv k_L^\mu x_\mu$ with $k_L^\mu = (\omega_L,\mathbf k_L)$
and $k_X^\mu =\varkappa k_L^\mu$.
We consider orthogonal linear polarizations, $A_X \cdot A_L=0$.
The vector potentials are parametrized as
$A_j^\mu (\phi )= \frac{m a_j}{e} \epsilon_j^\mu g_j(\phi) \cos \phi$,
$j \in \{X, L\}$,
with the modulus of the electron charge $e$,
normalized polarization vectors $\epsilon_j^\mu$ and
invariant laser strength parameters $a_j$.
We emphasize the appearance of the
envelope functions $g_j$ accounting for the finite pulse lengths $T_j$ of both
the X-ray and the laser pulses which are assumed to be synchronized temporally.

For the given background field $A^\mu$
we may work in the Furry picture employing Volkov states \cite{Volkov:1935,DiPiazza:RevModPhys2012}
\begin{align}
\Psi_p(x,A) &= e^{-ip\cdot x - \frac{i}{2k_L\cdot p} \int\limits^\phi \d \phi' [2 e p\cdot A(\phi') - e^2 A^2(\phi')]} 
 \nonumber \\
& \qquad \times \left( 1+\frac{e\slashed k_L \slashed A}{2 k_L \cdot p}\right) u_p \,, 
\end{align}
as non-perturbative solutions of the Dirac equation
$(i\slashed \partial - e \slashed A -m )\Psi_p(x)=0$.
The free Dirac spinors $u_p$ fulfil $(\slashed p-m)u_p=0$ and are normalized to
$\bar u_p u_p = 2m$. Moreover, Feynman's slash notation $\slashed p = \gamma_\mu p^\mu$ is employed.

The $S$ matrix for the emission of the photon $X'$ with momentum $k'$
is given by
\begin{align}
S& = -ie \int \! \d^4x \, \bar \Psi_{p'}(x,A) \slashed \epsilon' e^{ik'\cdot x} \Psi_p(x,A) \,,
\label{eq:S}
\end{align}
as depicted in Fig.~\ref{feyn:lacs}(a).
The non-perturbative expression \eqref{eq:S}, in both $a_X$ and $a_L$, is correct for arbitrary 
X-ray and laser intensities, describing multi-photon processes for both the laser and X-ray field.
For instance, in the papers \cite{Gush:PRD1975,Voroshilo:LasPhys1996,
Krajewska:PRA2012d,Augustin:PRA2013,Voroshilo:IEEE2013}
situations have been discussed where it is necessary to treat both field on equal footing.
However, even for present XFEL technology $a_X\ll 1$, and
we may expand the $S$ matrix into a power series in $a_X$
to get
\begin{align}
S =  S_0 + a_X (S_{+1} + S_{-1}) + \mathcal O(a_X^2)\,.
\label{eq:perturbation}
\end{align}
This corresponds to treating the field $A_X$ as a perturbation, going to a single X-ray photon
approximation.
A similar technique of expanding a Volkov wavefunction in a weak field has also been used, e.g.~in \cite{Herrmann:AnnPhys1972,Zhukovskii:JETP1973,book:Baier,Hu:PRL2011}.
Here, we first discuss the physical implications of that expansion
and give an interpretation for the individual terms of Eq.~\eqref{eq:perturbation}.
Details of the derivation are presented in the next section.

The lowest order term $S_0 \propto a_X^0$ (see Fig.~\ref{feyn:lacs}{(b)})
corresponds to non-linear
optical-laser Compton scattering without any participation of X-ray photon
\footnote{There are corrections to that process of the order of $a_X^2$ 
when arranging the expansion of Fig.~\ref{feyn:lacs}(a) according to the energy-momentum
balance $p + \ell k_L  = p' + k'$
\cite{Herrmann:AnnPhys1972}.}. 
This scattering of laser photons off free electrons is named commonly non-linear 
Thomson or Compton scattering,
see e.g.~\cite{Brown:PR1964,Narozhnyi:JETP1964,Narozhnyi:JETP83,Harvey:PRA2009,
Boca:PRA2009,Seipt:PRA2010,Seipt:PRA2011,Mackenroth:PRA2011}.
In the classical picture, this is the radiation due to the accelerated figure-8 motion of the electron in the laser. In the present context it may be dubbed spontaneous emission.
Using moderately strong laser pulses with $a_L\sim 1$,
the non-linear interaction with more than one photon
has been verified experimentally
\cite{Englert:PRA1983,Bula:PRL1996,Chen:Nature1998} via the observation of
harmonic radiation.
In these experiments, only a rather small
number of $\mathcal O(2\ldots5)$ laser photons was
participating in a single scattering event.

The leading order process involving X-ray photons is 
proportional to $a_X^1$ and consists of two terms which correspond to the absorption
($S_{+1}$, Fig.~\ref{feyn:lacs}(c))
or the emission  ($S_{-1}$, Fig.~\ref{feyn:lacs}(d))
of a single X-ray photon
from or to the initial beam.
The laser assisted Compton scattering is described by $S_{+1}$
with the formal energy-momentum conservation 
\begin{align}
p + k_X + \ell k_L  =  p' + k' \,, \label{eq:emc}
\end{align}
which arises upon the integration in \eqref{eq:S} together with introducing
the auxiliary variable $\ell$;
note that one still has to integrate over $\ell$, therefore \eqref{eq:emc} has
only three independent conservation laws for the light-front components \cite{Boca:PRA2009,Mackenroth:PRA2011}
\begin{align}
\mathbf p^\perp  &= \mathbf p'^\perp + \mathbf k'^\perp \,, &
p^- &= p'^-+k'^-\,.
\label{eq:emc_lc}
\end{align}
The light-front is defined with respect to $k_L$ such that $k_L^+=2\omega_L$
is the only non-vanishing light-front component of $k_L$
and $\phi  =\omega_L x^-$ becomes the light-front time-evolution parameter.
By adopting a coordinate system in which the laser propagates in the $z$-direction the light-front components of any four-vector read $x^\pm=x^0 \pm x^3$
and $\mathbf x^\perp = (x^1,x^2)^T$.

The fourth condition in \eqref{eq:emc}, $p^+ +k_X^+ +\ell k_L^+ = p'^+ + k'^+$, furnishes
a relation between the frequency $\omega'$ and the variable $\ell$ via Eq.~\eqref{eq:frequency}.
This may also be turned around expressing $\ell$ as a function of $k'$
via
\begin{align}
\ell(\omega',\vartheta) = \frac{k'\cdot p - k_X\cdot(p-k')}{k_L\cdot(p-k')}\,. \label{eq:L}
\end{align}
Thus, one can chose any of the two quantities
$\omega'$ or $\ell$ to be the independent variable defining the other one.
The lower limit of $\ell$ is determined by the condition
$\omega'\geq 0$, yielding $\ell \geq -\varkappa$.

The meaning of $\ell$ is that it parametrizes the amount of laser four-momentum $k_L$
absorbed in the scattering process
and it is the Fourier conjugate to the laser phase $\phi$ \cite{Heinzl:PLB2010,Harvey:PRL2012}. 
It can be considered as a continuous analogue of the photon number encountered for
infinite monochromatic plane wave fields \cite{Ilderton:PRL2011,Harvey:PRL2012},
cf.~also Appendix \ref{app:IPW}.

Due to the large
frequency ratio the two partial processes (b) and (c) are separated kinematically:
While photons from spontaneous emission $S_0$ have typical energies of $\omega' =\mathcal O(\electronvolt)$ (Eq.~\eqref{eq:frequency} with $\omega_X=0$),
the photons from laser assisted Compton scattering $S_{+1}$
are $\omega' = \mathcal O(\kilo \electronvolt)$.
The induced process $S_{-1}$, related to double Compton scattering \cite{Jentschura:PRL,Seipt:PRD2012,Mackenroth:PRL2013},
is strongly suppressed here in the $\kilo\electronvolt$ frequency
range due to the large value of $\varkappa$.

\section{Cross Section for Laser Assisted Compton Scattering}

\label{sect:crosssection}
Here we present the derivation of the matrix element for
laser assisted Compton scattering $S_{+1}$ which is needed to calculate the
corresponding cross section.
To this end
we have to linearise expression \eqref{eq:S} 
in $a_X$ and extract the part representing the absorption of an X-ray photon from the field $A_X$,
i.e.~a process with the formal energy momentum conservation \eqref{eq:emc}.
For these purposes it is convenient to split the integrand of $S$ in Eq.~\eqref{eq:S} into
the pre-exponential term
$\Gamma(x)=(1+ \frac{e\slashed A \slashed k_L}{2k_L\cdot p'}) \slashed \epsilon' 
(1+ \frac{e\slashed k_L \slashed A}{2k_L\cdot p}) $ 
and a phase factor
 $F(x)=(p'+k'-p)\cdot x 
 -  \int \! \d \phi \, ( \frac{e p\cdot A(\phi)}{k_L\cdot p} - \frac{e p'\cdot A(\phi)}{k_L\cdot p'} )
 +  ( \frac{e^2}{2k_L\cdot p} - \frac{e^2}{2k_L\cdot p'})  \int \! \d \phi  \, A^2(\phi)$,
such that
\begin{align}
S &= -ie \int \! \d^4x \, \bar u_{p'} \Gamma(x) u_p e^{i F(x)} \label{eq:S2} \,,
\end{align}
where
terms proportional to $a_X^2$ both in the exponent $F$ and the pre-exponential $\Gamma$
can be dropped by virtue of $a_X \ll 1$. We need to linearise furthermore
the phase exponent according to
\begin{align}
e^{iF} \approx e^{iF_0 + i F_X} \approx e^{iF_0} ( 1 + i F_X)\,,
\end{align}
where $F_0$ is the part of the phase of $S$ independent of $A_X$, i.e.~$\propto a_X^0$, and
\begin{align}
F_X &= \alpha_X \int \! \d(\varkappa \phi) \,  g_X(\varkappa\phi) \cos (\varkappa \phi) 
\end{align}
is linear in $a_X$ with
\begin{align}
\alpha_X &= m a_X \left( \frac{\epsilon_X\cdot p'}{k_X\cdot p'}  -  \frac{\epsilon_X\cdot p}{k_X\cdot p}\right)
\label{eq:alpha} \,.
\end{align}
Multiplying with the pre-exponential we obtain up to linear order in $a_X$
\begin{align}
\Gamma e^{iF} & 
\approx \Gamma_0 e^{iF_0} + (\Gamma_X + i F_X \Gamma_0) \,e^{iF_0} \,,
\label{eq:lin1}
\end{align}
where $\Gamma_0 = \sum_{n=0}^2 V_n^L [g_L(\phi) \cos \phi]^n$ is independent of $A_X$,
and $\Gamma_X = V_1^X g_X(k_X\cdot x) \cos k_X\cdot x $ is linear in $a_X$.
The pre-exponential coefficients are defined by
\begin{align}
V_0^j &= \slashed \epsilon' \,,  \label{eq:V0}\\
V_1^j &=  ma_j \left( \frac{\slashed \epsilon_j \slashed k_j \slashed \epsilon'}{2k_j\cdot p'}
         +  \frac{\slashed \epsilon' \slashed k_j \slashed \epsilon_j}{2k_j\cdot p} \right) \,, \\
V_2^j &=  \frac{m^2 a_j^2 \, \epsilon'\cdot k_j}{2 p\cdot k_j p'\cdot k_j} 
				\slashed k_j \,, \label{eq:V2}
\end{align}			
where $j\in (X,L)$ as above.
In expression \eqref{eq:lin1}, the first term gives rise to $S_0$ and the second term is $\propto a_X$ and contains both
$S_{\pm1}$, since the real field $A_X$ contains both the amplitudes for photon absorption and
emission.
Having linearised the expressions in $A_X$ we may now go over to
a complex field via $\cos (k_X\cdot x) = (e^{ik_X\cdot x} + e^{-ik_X\cdot x} )/2 \to e^{-ik_X\cdot x}/2$, selecting only the amplitude for
an X-ray photon in the initial channel,
yielding
\begin{align}
S_{+1} \propto \frac12 \left( 
				V_1^X g_X - \alpha_X \Gamma_0 G_X 
				\right)
				 e^{i F_0 -i k_X \cdot x}
\end{align}
with $G_X = -i e^{i\varkappa \phi} \int^{\varkappa \phi} \d \tau g_X(\tau) e^{-i\tau}$
and $\alpha_X$ defined in \eqref{eq:alpha}.
Employing the slowly varying envelope approximation (see e.g.~\cite{Seipt:PRA2011})
for the X-ray pulse means $G_X \to g_X$.
Physically, this approximation means neglecting a spatial displacement of the electron due to the
action of the X-ray field. The order of magnitude of this effect is given by $1/\omega_X T_X$, which is
much smaller than unity for femtosecond X-ray pulses of several $\kilo\electronvolt$ frequencies.

Performing the space-time integrations in \eqref{eq:S2},
the matrix element $S_{+1}$ for laser assisted Compton scattering can be written as
\begin{align}
S_{+1}= -4ie\pi^3 \int \! \d \ell \, \delta^{(4)}( p + \ell k_L + k_X - p'-k' )  M(\ell)
\label{eq:S+1}
\end{align}
with the amplitude
\begin{align}
M(\ell) &= \bar u_{p'} 
\left[
V_1^X C_0(\ell)  - \alpha_X \sum_{n=0}^2  V_n^L C_n(\ell)
\right]  u_p \,.
\label{eq:M}
\end{align}
and the coefficients $V_n^j$ defined in \eqref{eq:V0}--\eqref{eq:V2}.
The functions $C_n$ in \eqref{eq:M} are given by
\begin{align}
C_n 
  = \int _{-\infty}^\infty \! \d\phi \, [g_L(\phi)\cos\phi]^n 
g_X (\varkappa \phi)
e^{i\int^\phi \d\phi' \psi(\phi')}
\label{eq:Cfunction}
\end{align}
with the dynamic phase
\begin{align}
\psi(\phi) = \ell + \alpha_L g_L(\phi) \cos \phi + \beta_L [g_L(\phi)\cos\phi]^2
\label{eq:dynamical_phase}
\end{align}
having defined 
\begin{align}
\alpha_L &= m a_L \left( \frac{\epsilon_L\cdot p'}{k_L\cdot p'}
	-  \frac{\epsilon_L\cdot p}{k_L\cdot p} \right)
\intertext{and}
\beta_L &= \frac{m^2 a_L^2}{2} \left( \frac{1}{ k_L \cdot p' } - \frac{1}{ k_L \cdot p } \right)  \,.
\end{align}
Upon performing the integration over $\d\ell$ in \eqref{eq:S+1}
the argument of the amplitude $M$ becomes
a function of $\omega'$ and $\vartheta$, i.e.~$M(\ell(\omega',\vartheta))$,
with $\ell(\omega',\vartheta)$ given by Eq.~\eqref{eq:L}.

The cross section
\begin{align}
\frac{\d \sigma}{ \d \Omega \d \omega'} & = \frac{r_e^2}{8\pi \int_{-\infty}^\infty \!  \d \phi \, g_X(\phi)^2} \frac{  \omega' \left| M(\ell(\omega',\vartheta)) \right|^2 }{ ( k_L\cdot p)( k_L\cdot p')} ,
\label{eq:crosssection}
\end{align}
depending on spin and polarization variables, is normalized to the incident X-ray flux
($r_e \simeq \unit{2.8}{\femto \metre}$ is the classical electron radius).
Equation~\eqref{eq:crosssection} is differential in the final photon momenta.
The final state electron is supposed to remain unobserved, i.e.~the final electron momentum $p'$ is integrated out and its value is fixed by Eq.~\eqref{eq:emc_lc}.
In the limit $a_L\to0$, i.e.~a vanishing laser field, we recover
from \eqref{eq:M}
the KN matrix element 
and from Eq.~\eqref{eq:crosssection} the KN cross section \cite{Klein:ZPhys1929},
however, both ones with the initial photon $X$ described as a wave packet.

We emphasize the finite X-ray pulse length $T_X$ encoded in $g_X$.
Without $g_X$,
the integral $C_0$ would
diverge as $1/\ell$ for $\ell\to 0$,
even after a regularization similar to \cite{Boca:PRA2009}.
This is no issue for non-linear Compton scattering
since there $\ell=0$ implies $\omega'=0$ \cite{Seipt:PRA2011}.
Here, however, $\ell=0$ denotes the KN line.
The behaviour of $C_0$ at $\ell=0$ can be related to the scattering of X-ray photons during
the time interval outside the laser pulse, where no laser photons are exchanged.
The time integrated probability for that partial process
grows $\propto T_X$ for large $T_X$
while the probability for the scattering inside the laser pulse
stays finite for finite $T_L$.
This leads to a relative suppression of the influence of the laser pulse for large values of $T_X$.
In \cite{Nedoreshta:LasPhys2013}, where laser assisted Compton scattering
was studied in a pulsed laser field combined with infinite monochromatic X-ray waves,
the authors introduced a short finite observation time
to fix this issue.
In our approach, working with finite X-ray pulses from the beginning,
we naturally obtain consistent results,
where the value of $T_X$ is related to the specific experimental conditions.

\section{Properties of the Spectrum}
\label{sect:spectrum}

In the following we calculate the spectrum of laser assisted Compton scattering numerically for a representative choice of parameters to exhibit the essential features.

\subsection{Choice of Parameters}

The following numerical results are for
an experiment which could be realized when combining an XFEL 
(e.g.~LCLS or  European XFEL) with an optical laser.
That means, despite of the coherence properties, the XFELs are considered as
sources of short, almost monochromatic X-ray photon pulses. 
For the sake of definiteness we specify
$\omega_X =\unit{5}{\kilo\electronvolt}$ for the X-rays as well as an
$\unit{800}{\nano\metre}$
Ti:Sapphire laser, i.e.~$\omega_L= \unit{1.55}{\electronvolt}$.
The planned Helmholtz international beamline for extreme fields (HIBEF) \cite{site:HIBEF} 
at the European XFEL \cite{XFEL} will provide such a set-up. 
We envisage laser intensities $I\sim\unit{10^{18}}{\watt\per\centi\metre^2}$,
where
$a_L = 0.68 \sqrt{I [\unit{10^{18}}{\watt\per\centi\metre^2}]}$. 
The optical laser pulse length is set to
$T_L=\unit{20}{\femto\second}$
and the X-ray pulse length is taken as
$T_X= \unit{50}{\femto\second}$ (both FWHM values),
in agreement with routinely achieved optical laser pulses and XFEL design
\cite{XFEL}.
For convenience we use a $\cos^2$ profile \cite{Seipt:PRD2012} for $g_L$ and a Gaussian for $g_X$.

In the following we calculate the photon spectra for laser assisted Compton scattering in the rest frame of the initial electron and
focus on photon energies $\omega'= \mathcal O(\kilo\electronvolt)$ in
the vicinity of the KN line.

In an actual experiment one could employ low-energy electrons emitted from an electron gun \cite{Englert:PRA1983}.
The results from the electron rest frame can be boosted to the laboratory frame using
Lorentz transformations. For low electron kinetic energies $E_{\rm kin}\ll m$, e.g.~the
frequencies are transformed as
$\omega'_{\rm lab} \simeq \omega' (1 + \frac{v}{2} (1 + \cos \vartheta))$,
where $v = \sqrt{2E_{\rm kin}/m}$ and scattering angles close to
$\vartheta=\pi/2$ transform as
$ \vartheta_{\rm lab} \simeq \vartheta + v$.

\subsection{Energy spectra}

\begin{figure}
\begin{center}
\includegraphics[width=0.49\columnwidth]{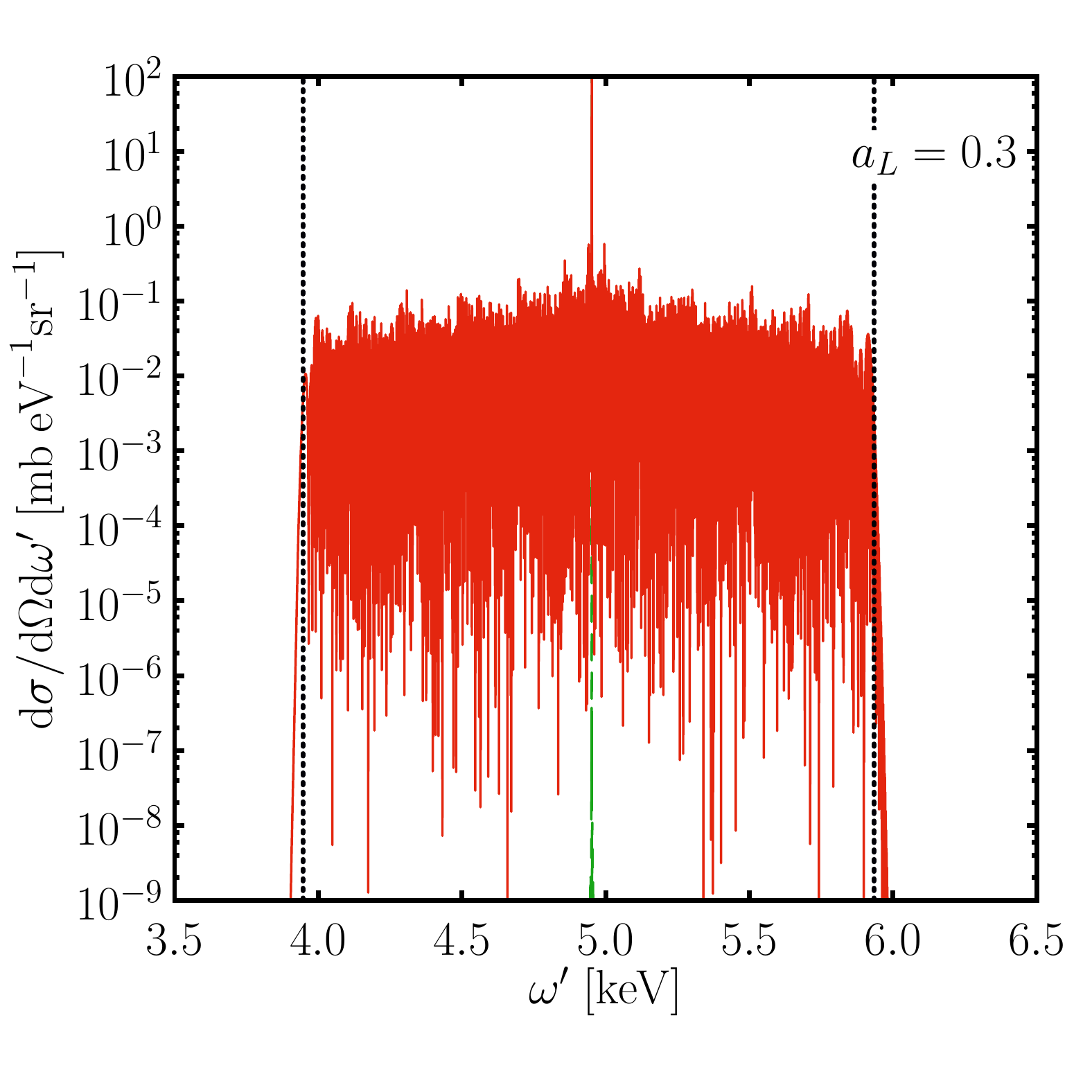}
\includegraphics[width=0.49\columnwidth]{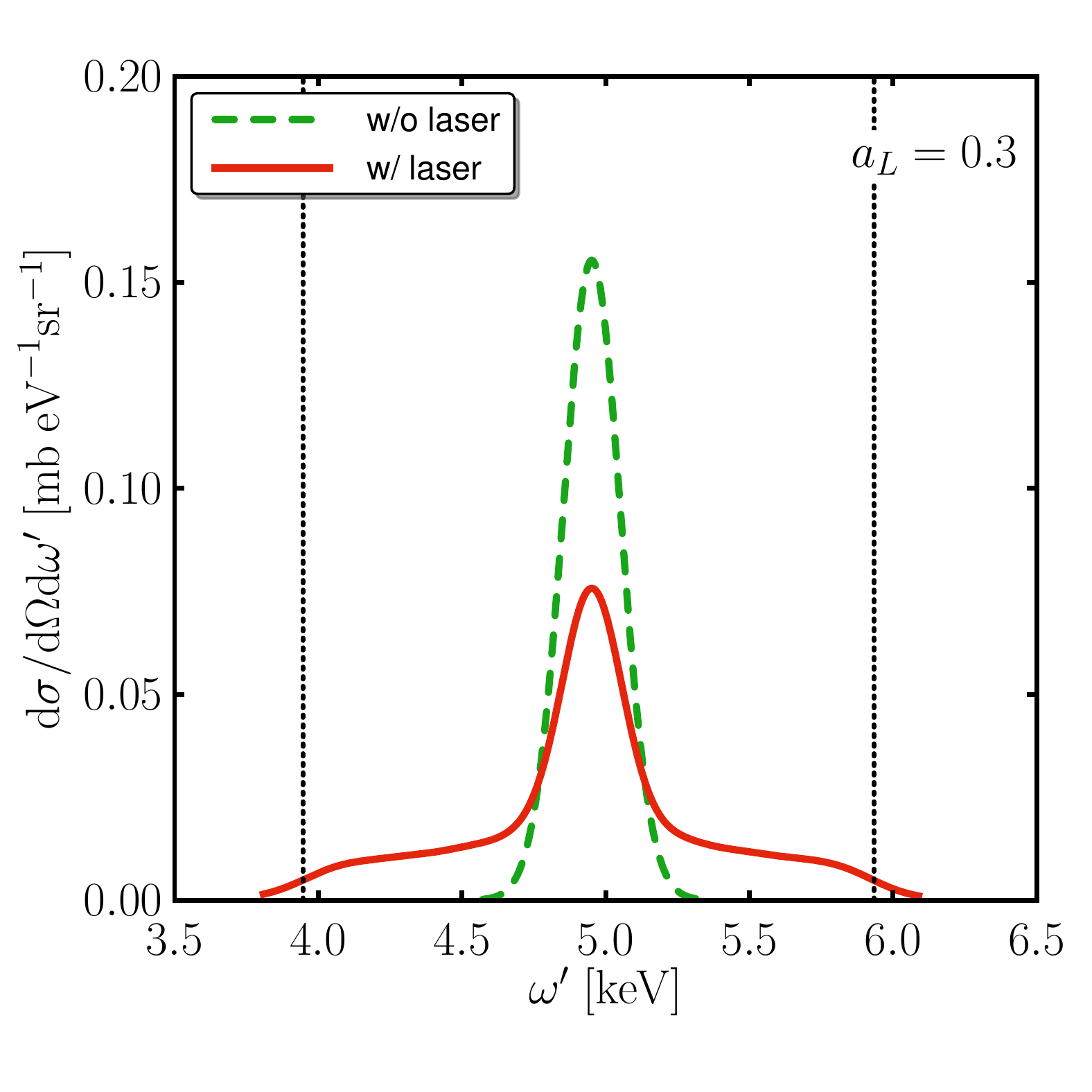}
\end{center}
\caption{(Color online)
Differential cross section
(left panel in a log scale) 
as a function of $\omega'$ for the fixed observation direction
$\vartheta = 90\degree$ and $\varphi=45\degree$ for 
$a_L=0.3$. The vertical dotted lines depict the cut-off values (\ref{eq:cutoff1},\ref{eq:cutoff2}).
The right panel shows, in a linear scale, the
spectrum with $\unit{100}{\electronvolt}$  (rms Gaussian shape) resolution
(red solid curve; for a comparison, the plain KN line without the laser is depicted as green dashed curve.
}
\label{fig:spectrum_omega}
\end{figure}

The unpolarized differential cross section
is exhibited in Fig.~\ref{fig:spectrum_omega}
as a function of the frequency $\omega'$
for a fixed observation direction
$\vartheta = 90\degree$ and $\varphi=45\degree$.
The scattering angle $\vartheta$ is measured with respect to\ the
beam axis $\mathbf k_L$;
the azimuthal angle $\varphi$ w.r.t.\ $\mathbf \epsilon_L$.
In the left panel of Fig.~\ref{fig:spectrum_omega}, for $a_L=0.3$
many side-lines
form a roughened plateau, where the main KN line sticks out
being very narrow as compared to the width of the plateau.
The
plateau spans the range
$\unit{4\ldots6}{\kilo\electronvolt}$, corresponding to
the order of $\unit{1}{ \kilo\electronvolt}/\omega_L\sim 650$ exchanged laser photons, despite of $a_L <1$.
The right panel depicts the spectrum, averaged
with a detector resolution of $\unit{100}{\electronvolt}$, on a linear scale.
While the KN line becomes broader, the plateau is clearly visible
and the roughness is averaged out (red curve).
Compared to Compton scattering without the laser (green dashed curve),
the peak height of the Klein Nishina line is reduced to half of its value
(see also the discussion in Section \ref{subsect:D})

To find the relevant multi-photon parameter for the process in Fig.~\ref{feyn:lacs}(c)
we determine 
the cut-off values $\ell_{\pm}$, beyond which the
amplitude drops exponentially fast,
via the stationary phase method by requiring that
there are no stationary points $\phi_\star$ on the real axis,
i.e.~$\psi(\phi_\star)=0$ (see Eq.~\eqref{eq:dynamical_phase})
has no real solutions.
Abbreviating $f = g_L(\phi) \cos \phi$ we find
\begin{align}
f(\phi_\star) = - \frac{\alpha_L}{2\beta_L} \pm \sqrt{\frac{\alpha_L^2}{4\beta_L^2} - \frac{\ell}{\beta_L} } \,,
\end{align}
which has no real solutions for $\phi_\star$ if $(i)$ the term under the root
becomes negative or $(ii)$ there are no real solutions when calculating the inverse function $ f^{-1}$.
The cut-off values are determined at maximum intensity at the center of the pulse, thus, we may set $g_L\to 1$ here and
condition $(ii)$ turns into $|f| \leq 1$.
We find
\begin{align}
\ell_{-} &=   \frac{\varkappa \xi_-}{1-\xi_-}  \,, 
\label{eq:cutoff1} \\
\ell_{+} & = 
\left\{
\begin{matrix}
		\displaystyle  \frac{\varkappa \xi_+}{1 - \xi_+}\,,  & 
	\quad	\text{for }	\vartheta \leq \vartheta_\star \,, \\
		\displaystyle \frac{\varkappa \xi_0}{1-\xi_0} 	\,, 	& 
		\quad	\text{for }	\vartheta > \vartheta_\star
\end{matrix} \right.
\label{eq:cutoff2}
\end{align}
with 
\begin{align}
\xi_0 &=\frac{\sin^2\vartheta \cos^2\varphi}{2-2\cos\vartheta}\,, \\
\xi_\pm &=  \frac{a_L^2}{2}(\cos \vartheta - 1) 
\pm a_L \sin \vartheta |\cos \varphi| \\
\intertext{and}
\cos \vartheta_\star &= \frac{a_L^2-\cos^2\varphi}{a_L^2 + \cos^2\varphi}\,.
\end{align}
For $a_L < 1$, the cut-off values are of the order $\varkappa a_L$ for most
observation angles. 
Thus, for moderately strong laser fields
and a large frequency ratio the relevant multi-photon parameter is
$\varkappa a_L$.
This is in severe contrast to the spontaneous process $S_0$ depicted in Fig.~\ref{feyn:lacs}(b), where the multi-photon parameter is $a_L$ \cite{Ritus:JSLR1985}.

\subsection{Angular spectra}

\begin{figure}
\includegraphics[width=0.5\columnwidth]{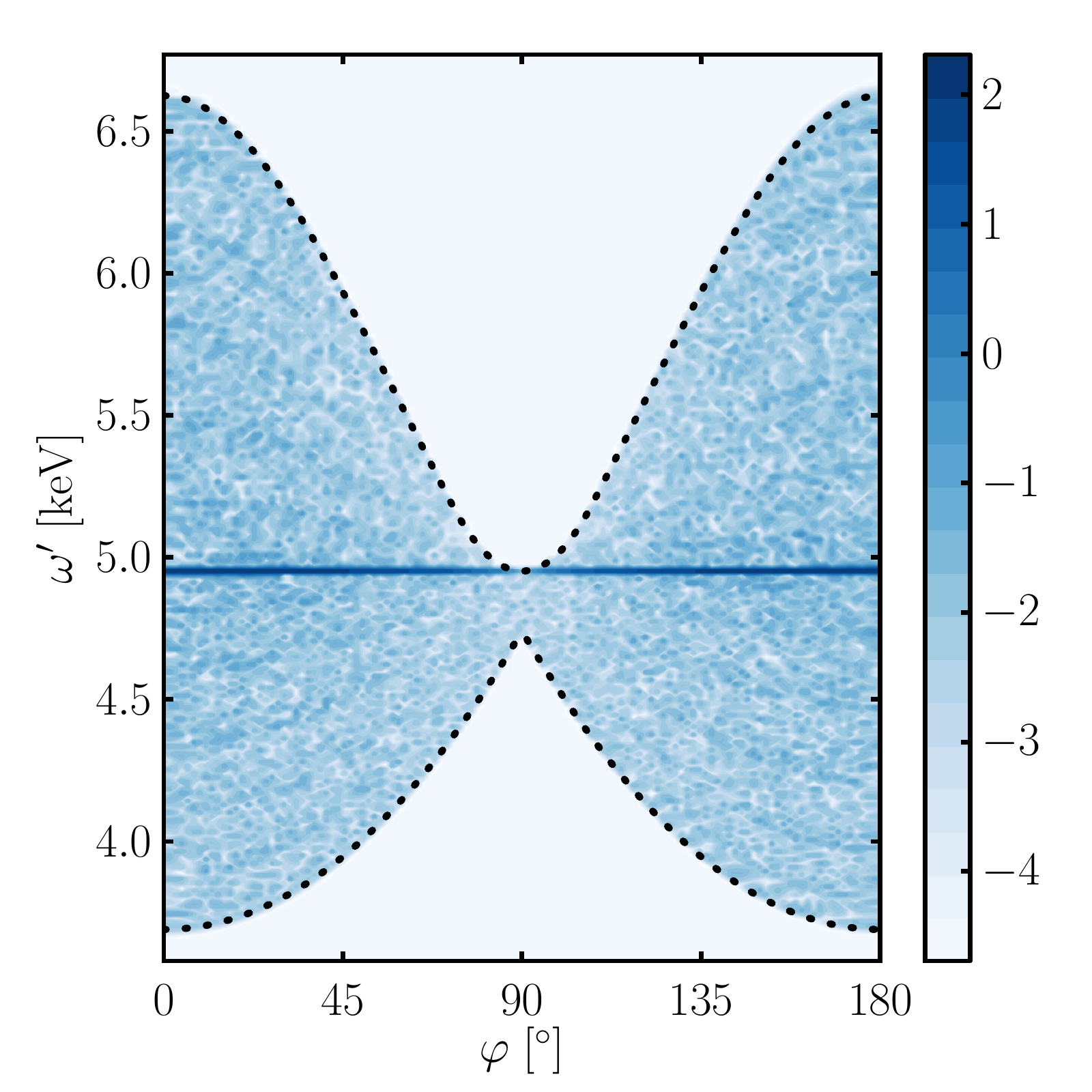}
\hspace*{-0.03\columnwidth}
\includegraphics[width=0.5\columnwidth]{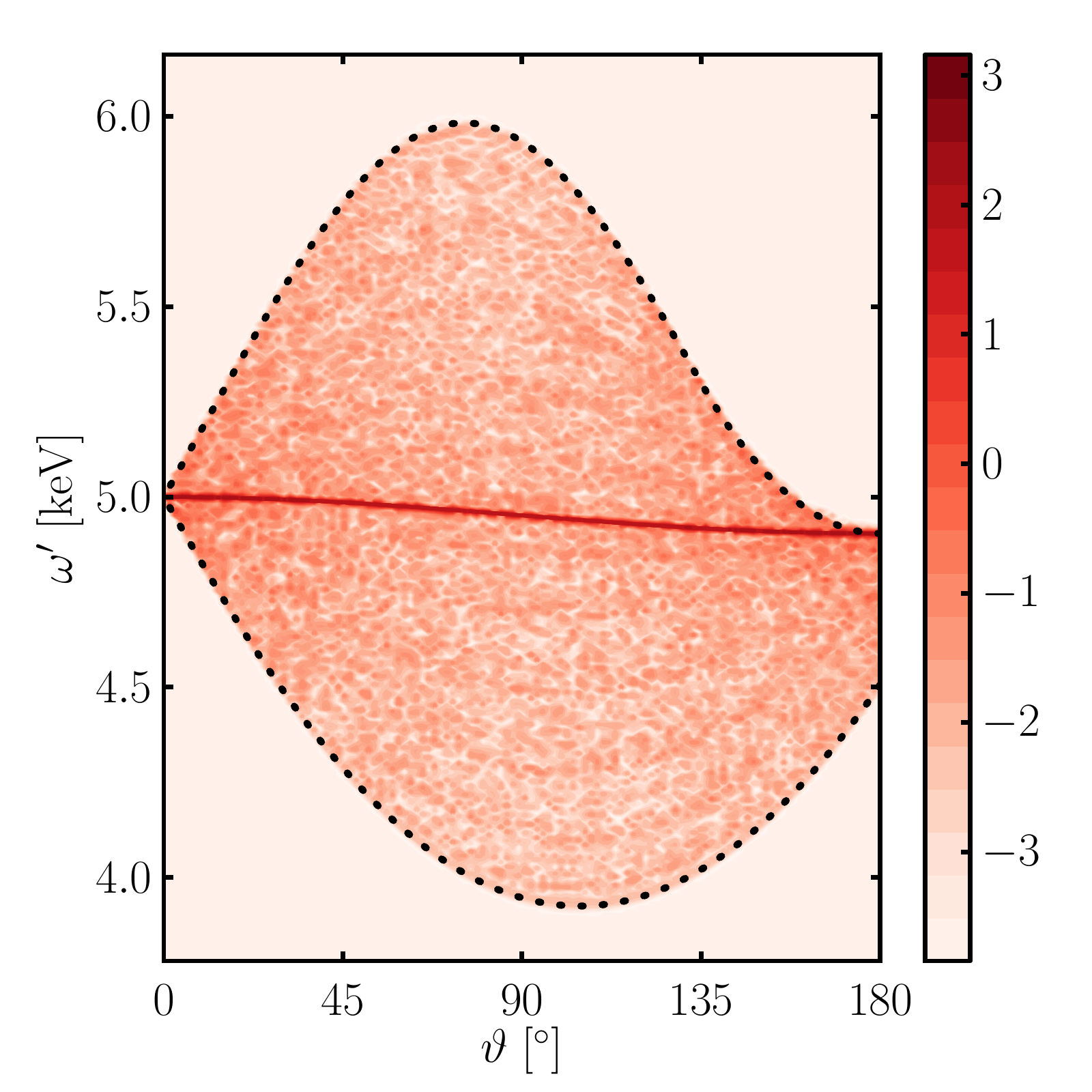}
\caption{(Color online)
Frequency vs. angle correlations of the cross section
as a function of the
azimuthal angle for fixed $\vartheta=90\degree$ (left panel) and
as a function of $\vartheta$ for fixed $\varphi=45\degree$ (right panel) for $a_L=0.3$.
The cutoff-values (\ref{eq:cutoff1},\ref{eq:cutoff2}) are
depicted by the dotted curves.
The color scale represents
$\log_{10} ( 
\d \sigma / \d \Omega \d \omega' \usk 
[\milli\barn\usk \electronvolt^{-1}\steradian^{-1} ])$.
}
\label{fig:spectrum}
\end{figure}

The plateau has a pronounced dependence on the azimuthal angle $\varphi$,
see left panel of Fig.~\ref{fig:spectrum}. A strong reduction of the plateau width is observed at $\varphi=90\degree$, i.e.~perpendicularly to the laser polarization,
where $\ell_+=0$.
The broadest plateau is found close to the laser polarization direction
which is perpendicular to the X-ray polarization where up to $1000$ laser photons
participate.
The polar angle distribution (right panel) shows a forward-backward asymmetry.
For large scattering angles $\vartheta > 90\degree$ the emission with $\ell>0$,
i.e.~$\omega'>\omega'_{\rm KN}$ is suppressed.
The cut-off values (\ref{eq:cutoff1},\ref{eq:cutoff2}),
depicted by dotted curves, coincide with the numerical results.
From our analysis of the angular spectra we propose to
choose as observation direction $\vartheta = 90\degree$ and
an azimuthal angle not too close to $\varphi=90\degree$, e.g.~as
discussed above in Fig.~\ref{fig:spectrum_omega},
to optimize the width of the plateau.

\subsection{Energy Integrated Cross Section}

\label{subsect:D}
Numerically we find that the energy integrated cross section
equals the KN cross section \cite{book:Landau4}, i.e.\
\begin{align}
\int \! \d \omega' \,\frac{\d \sigma}{\d\Omega\d \omega'}  = \frac{d\sigma_{\rm KN}}{\d\Omega}  \,. 
\end{align}
A qualitative argument for this behavior is provided by a classical model,
where the total emitted power $\dot E$ is given by the Larmor formula yielding
\begin{align}
\dot E &= 
- \frac{e^4}{6\pi m^2} ( A'_L\cdot A'_L + A'_X\cdot A'_X) \,,
\end{align}
where primes denote derivatives w.r.t.~$\phi$.
The first term in brackets corresponds to the spontaneous
emission process and the second term refers to
the laser assisted Compton scattering of X-ray photons $S_{+1}$.
The latter part is independent of the laser intensity.
Such a redistribution in phase space with
marginal impact on the total probability
has been found also
for other laser assisted processes \cite{Loetstedt:2008}.

To quantify the fraction of photons scattered into the side-bands,
in particular for the observation direction used in Fig.~\ref{fig:spectrum_omega},
we define the side-band cross section as that part of the spectrum which is at least by $\pm \omega_L/2$ away from $\omega'_{\rm KN}$.
(A variation of the discrimination value
in the range $0.25\ldots0.75\,\omega_L$ leads to
a relative uncertainty of $ \le 5\%$.)
For small values of $\varkappa a_L \ll 1$, the side-band fraction scales as $a_L^2$.
For larger values of $a_L$, the main line is weakened,
see right panel in Fig.~\ref{fig:spectrum_omega}. 
(For monochromatic waves, where $g_{X, L} \to 1$, 
also a negative lowest-order $a_L$ corrections to the main line has been found  \cite{Gush:PRD1975,Akhiezer:JETP1985}).
The side-band cross section increases with $a_L$ up to
$a_L \sim 0.01$, where it saturates at $\unit{30}{\milli\barn \usk \steradian^{-1}}$.
This value should be
compared to the corresponding KN cross section (without the laser) of
$\unit{39}{\milli\barn \usk \steradian^{-1}}$,
proving that a large fraction of the photons is emitted into the side-bands.

\subsection{Polarization}

\begin{figure}
\center~\includegraphics[width=0.99\columnwidth]{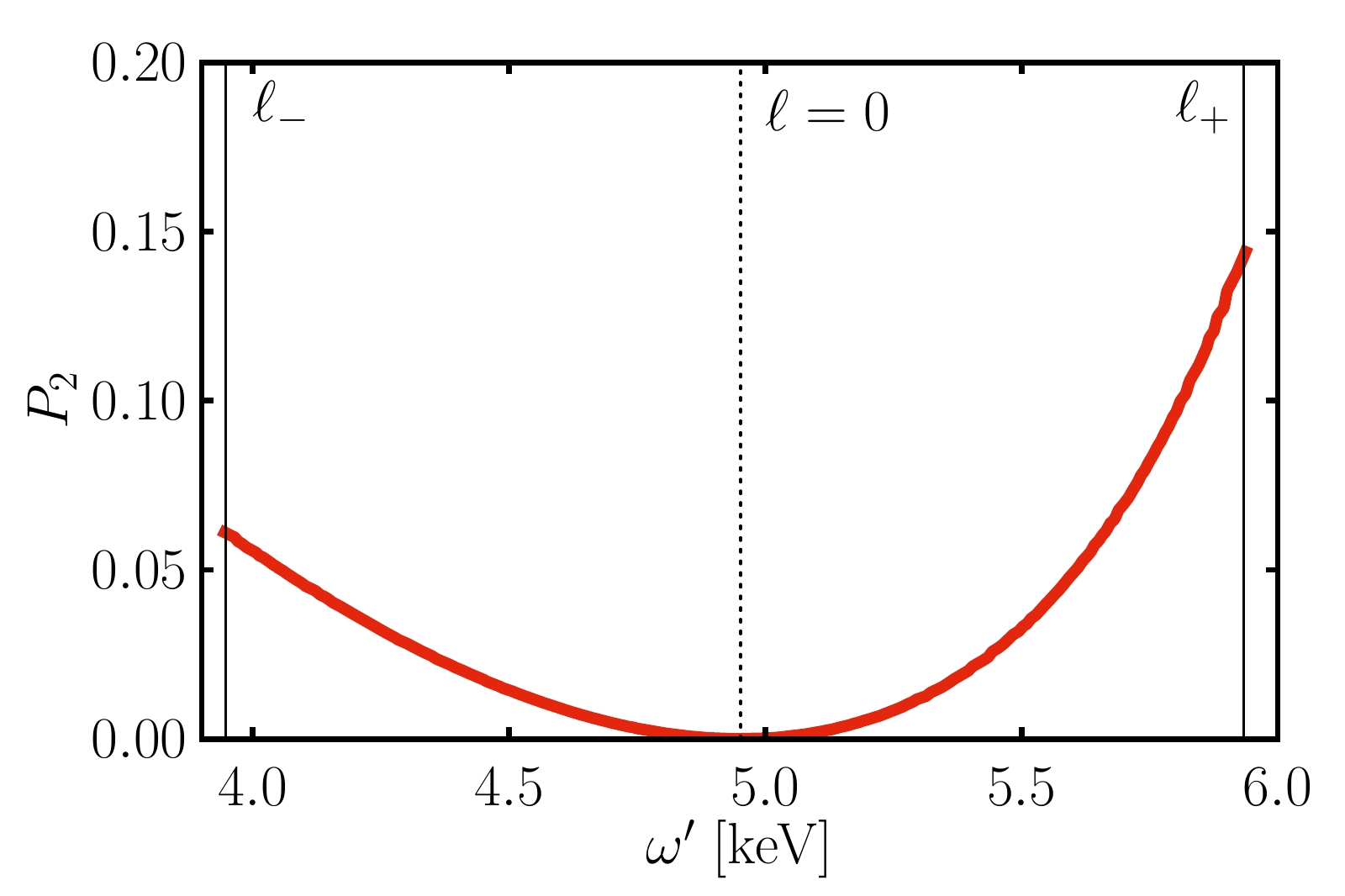}
\caption{(Color online)
Rotation of the polarization of the scattered photon, quantified via
the Stokes parameter $P_2$ as a function of $\omega'$ for
the same observation direction as in Fig.~\ref{fig:spectrum_omega}.
}
\label{fig:polarization}
\end{figure}

Due to the non-linear mixing of optical laser photons with the incident X-ray photon
the polarization of the photon $X'$ is rotated
as compared to Compton scattering without the laser.
The rotation angle is frequency dependent.
This rotation can be quantified experimentally by measuring the Stokes parameter $P_2$ which we calculate as
$P_2 = (w_{45\degree} - w_{135\degree}) /
		   ( w_{45\degree} + w_{135\degree}  )$, where
$w_\chi$ denotes the triple differential
cross section of photons with their polarization vector having an angle of $\chi$ w.r.t.~the scattering plane.
For instance, for the observation angles of Fig.~\ref{fig:spectrum_omega}
the photon $X'$ is polarized perpendicular to the scattering plane without the
laser which means $P_2=0$.
This result is changed in laser assisted Compton scattering.
In Fig.~\ref{fig:polarization} the Stokes parameter $P_2$ is exhibited as a function of $\omega'$ for $a_L=0.3$. The value of $P_2$ is zero at the KN line
and increases with increasing distance from it, i.e., in the region  where more
laser photons are involved.
The maximum value of $P_2$ corresponds to a rotation angle of $5\degree$
towards $\mathbf k_L$.
In the classical picture,
the figure-8 orbit in the laser field has a component
in the directions of $\mathbf k_L$ which is transferred to the polarization of the photon $X'$.
The energy dependent polarization rotation can be used to identify unambiguously the final state photons emerging from the multi-photon process.

\section{Discussion and Summary}
\label{sect:summary}
A clear and easily observable signature of 
laser assisted Compton scattering of X-ray photons
is provided by  the side-lines accompanying the main
Klein-Nishina line forming a broad plateau. 
Due to the difference of scales of the photon energies
of X-ray and optical laser, 
non-linear effects are strongly enhanced as compared
to the spontaneous emission of radiation in a pure laser field.
The relevant multi-photon parameter is $\varkappa a_L$, which can
be large even for laser fields with intensities of the order of
$\unit{10^{18}}{\watt\per\centi\metre^2}$.
The optimal conditions to observe the side-band plateau 
with an X-ray camera are achieved for 
$a_L\lesssim 1$.
Such intensities are achieved routinely with a $\unit{200}{\tera\watt}$ laser
in relatively large spot sizes of $w_0=\mathcal O(\unit{100}{\micro\metre})$
and pulse lengths of $T_L=\unit{20}{\femto\second}$.
We do not expect that spatially inhomogeneous
laser spots influence the cut-off values since they are sensitive to
the maximum laser intensity in the spatio-temporal profile of the pulse. 
However, the shape of the plateau will certainly change.
The optical photons emerging from spontaneous emission process,
i.e.\ the non-linear Compton process exhibited in Fig.~\ref{feyn:lacs}(b),
can be efficiently filtered out by a thin foil which is otherwise 
transparent for X-rays.

A very useful signal for the non-linear frequency mixing is the frequency dependent
rotation of the polarization of the final state photons.
The polarization can be measured by using X-ray polarizers. A polarization purity of the order of $10^{-10}$ has been achieved recently \cite{Marx:PRL2013}.
The technique of a polarization veto can be used to shield the plain KN photons
(arising for unsynchronized X and L pulses) or shield the $\ell = 0$ KN-like
photons.
For optimal conditions, the X-ray pulse length $T_X$ and the laser pulse length $T_L$ should be of the same order of magnitude.
The synchronization of optical and X-ray pulses to the level of
femtoseconds has been achieved experimentally \cite{Tavella:NatPhot2011},
approaching the sub-femtosecond level \cite{Schultze:Science2010}.
Numerically, the spectra are insensitive to a temporal offset between the two pulses of the order of a few femtoseconds.

A slight misalignment of the X-ray and laser beams due to
axis off-sets or focusing effects does not lead to qualitatively new effects:
The Oleinik resonances \cite{Oleinik:JETP1968}
for non-parallel beams 
(see e.g.~\cite{diss:Hartin,Voroshilo:LasPhysLett2005,Voroshilo:LasPhys2011,Nedoreshta:PRA2013})
are suppressed by the large frequency ratio $\varkappa$ and
for a small angular misalignment $\Theta\ll1$ between the X-ray and laser beams.
A slight misalignment of the polarizations,
$\mathbf \epsilon_X \cdot \mathbf \epsilon_L \ne 0$,
causes laser intensity dependent modifications of the total cross section
of the order ${\cal O} (\mathbf \epsilon_X \cdot \mathbf \epsilon_L ) \ll 1$.

Such experiments can be performed at LCLS or 
at the European XFEL combining the X-ray facility with an optical laser system, 
e.g., as planned in HIBEF \cite{site:HIBEF}. 
One could employ low-energy electrons emitted from an electron gun
as in the experiment \cite{Englert:PRA1983}.
However, the electron energy cannot be too low since low energy-electrons are expelled from
high-intensity regions due to the ponderomotive force.
We found numerically that for electron kinetic energies as low as
$E_{\rm kin}=\unit{500}{\electronvolt}$ the electrons can penetrate the focus and the
maximum deflection angle is below $\unit{0.8}{\degree}$.
This is due to the fact that the transverse ponderomotive force which is responsible for the deflection
is proportional to the gradient of the laser intensity and
scales as $\propto a_L^2/w_0$, where $a_L$ is rather small and the laser spot size $w_0$ is large.
Consequently, electrons with kinetic energies of a few hundred $\electronvolt$
are suitable for such experiments.

In conclusion, studying the laser assisted Compton scattering of X-rays
will significantly advance our understanding of strong-field QED
scattering processes in the ${\cal O} (10^3)$ multi-photon regime.

\

\section{Acknowledgements}

The authors are grateful to T.~E.~Cowan and R.~Sauerbrey for the inspiring
collaboration on future experiments at HIBEF.
DS acknowledges stimulating discussions with
S.~Fritzsche, A.~Surzhykov and V.~G.~Serbo.

\appendix

\section{The limit of infinite monochromatic plane waves}
\label{app:IPW}

In this appendix we discuss briefly the limit of infinite monochromatic plane waves (IPW), $T_X,T_L\to \infty$.
In this case, characterized by $g_{L,X} \to 1$, the integral in the exponents of
the functions $C_n$, Eq.~\eqref{eq:Cfunction}, is given by
\begin{align}
\int \! \d \phi \, \psi_{\rm IPW}(\phi) &= \ell \phi + \alpha_L \sin \phi +\frac{\beta_L}{2} \phi + \frac{\beta_L}{4} \sin 2\phi \,.
\label{eq:psiIPW}
\end{align}
Owing to the periodicity of the monochromatic field,
the integrand of the $C_n$
can be expanded in a discrete Fourier series, excluding the non-periodic terms in \eqref{eq:psiIPW}, such
that
the integrals $C_n$ turn into a sum
over discrete partial amplitudes.
For instance, for $C_0$ we find 
\begin{align}
C_0^{\rm IPW} = \sum_{N=-\infty}^\infty 2\pi \delta( \ell - N + \beta_L/2 ) J_N(\alpha_L,\beta_L)
\label{eq:A2}
\end{align}
with amplitudes
\begin{align}
J_N(\alpha_L,\beta_L) = \sum_{s=-\infty}^\infty J_{N-2s}(-\alpha_L) J_{s}(-\beta_L/4)
\end{align}
and the Bessel functions $J_N$.
Similar relations can be found for all integrals $C_n^{\rm IPW}$.
Thus, in the limit of infinite plane waves the variable $\ell$ becomes a
discrete $\ell_N = N - \beta_L/2$, with integer values of $N$, due to $\delta(\ell - N + \beta_L/2)$.
The term $\beta_L/2$ can be
absorbed into the electron momenta leading to
the occurrence of the field-dressed quasi-momentum
\begin{align}
\tilde p = p + \frac{m^2 a_L^2}{4k_L\cdot p} k_L\,,
\end{align}
and similarly for $p'$.
The formal energy momentum conservation \eqref{eq:emc} turns into
\begin{align}
\tilde p + k_X + N k_L = \tilde p' + k' \,,
\label{eq:emcIPW}
\end{align}
where $N$ can be considered as number of laser photons exchanged during the scattering process,
as pointed out e.g.~in \cite{Mackenroth:PRA2011,Harvey:PRL2012} for non-linear Compton scattering.
Solving \eqref{eq:emcIPW} for the frequency $\omega'$
gives
\begin{align}
\omega'_N(\vartheta) = \frac{\omega_X + N \omega_L}
{1 + \left( \frac{\omega_X+N\omega_L}{m} + \frac{a_L^2}{4}\right) (1-\cos\vartheta)} \,,
\end{align}
i.e.~discrete frequencies $\omega'_N$
with an intensity dependent red-shift quantified by the term $a_L^2/4$ in the
denominator \cite{Ehlotzky:JPB1989}.

Integrating over $\d\ell$ in \eqref{eq:S+1} and exploiting the delta distributions
in the functions $C_n^{\rm IPW}$, the amplitude $M$, Eq.~\eqref{eq:M}, becomes a sum of discrete
partial amplitudes
\begin{align}
M_{\rm IPW} = \sum_N M_N \,.
\end{align}
The same is true for the cross section
\begin{align}
\frac{\d \sigma_{\rm IPW}}{\d \Omega} = \sum_{N} \frac{\d \sigma_{N}}{\d \Omega} \,,
\end{align}
as known from the literature, e.g.~\cite{Gush:PRD1975,Akhiezer:JETP1985,Ehlotzky:JPB1989}.
The dependence on $\omega'$ in \eqref{eq:crosssection} is replaced by a sum over discrete values
of $N$.
This shows that our results for pulsed plane waves, in particular \eqref{eq:S+1}, \eqref{eq:M}
and \eqref{eq:crosssection}, contain also
the known case of infinite plane waves in the proper limit.

\end{document}